\documentclass[12pt]{article}
\usepackage{amsmath}
\usepackage{graphicx}
\usepackage{geometry}
\usepackage[applemac]{inputenc}

\usepackage{natbib}

\ifx\pdfoutput\@undefined\usepackage[usenames,dvips]{color}
\else\usepackage[usenames,dvipsnames]{color}
\IfFileExists{pdfcolmk.sty}{\usepackage{pdfcolmk}}{} 
\fi
\usepackage[plainpages=false,pdfpagelabels,pagebackref=false,naturalnames=true,hyperindex=true,pdftitle={When slower is faster?},pdfauthor={}]{hyperref}
\hypersetup{colorlinks=true,
urlcolor=Cerulean,linkcolor=BrickRed,citecolor=RoyalBlue,
  pdfpagemode=None,
  pdfstartview=FitH}
\usepackage[all]{hypcap}

\begin{document}

\title{When slower is faster}
\author{Carlos Gershenson$^{1,2,3,4,5}$ and Dirk Helbing$^{6}$\\
$^{1}$
Instituto de Investigaciones en Matem\'aticas Aplicadas y en Sistemas, \\
Universidad Nacional Aut\'onoma de M\'exico.\\
\href{mailto:cgg@unam.mx}{cgg@unam.mx} \
\url{http://turing.iimas.unam.mx/~cgg} \\
$^{2}$ Centro de Ciencias de la Complejidad, UNAM, M\'exico. \\
$^{3}$ SENSEable City Lab, Massachusetts Institute of Technology, USA.\\
$^{4}$ MoBS Lab, Northeastern University, USA.\\
$^{5}$ ITMO University, St. Petersburg, Russian Federation.\\
$^{6}$Department of Humanities, Social and Political Sciences (GESS)\\
ETH Z\"urich \\
\href{mailto:dirk.helbing@gess.ethz.ch}{dirk.helbing@gess.ethz.ch} \
\url{http://www.coss.ethz.ch}
}
\maketitle

\begin{abstract}
The slower is faster (SIF) effect occurs when a system performs worse as its components try to do better. Thus, a moderate individual efficiency actually leads to a better systemic performance.
The SIF effect takes place in a variety of phenomena. We review studies and examples of the SIF effect in pedestrian dynamics, vehicle traffic, traffic light control, logistics, public transport, social dynamics, ecological systems, and adaptation. Drawing on these examples, we generalize common features of the SIF effect and suggest possible future lines of research.
\end{abstract}

\section{Introduction}

How fast should an athlete run a race? If she goes too fast, she will burn out and become tired before finishing. If she runs conservatively, she will not get tired, but will not make her best time. To minimize her race time, she has to go as fast as possible but without burning out. If she goes faster, she will actually race more slowly. This is an example of the ``slower-is-faster" (SIF) effect: in order to run faster, sometimes it is necessary to run slower, not to burn out. It is not trivial to calculate the running speed which will lead to the best race, as this depends on the athlete, race distance, track, temperature, humidity, and daily performance: Running 100m dash should be done as fast as you can, while running a marathon demands a carefully paced race. How fast would an athlete run a marathon if she started with a speed for a 100m? To finish the marathon successfully, she would obviously have to run more slowly.

There are several other examples of the SIF effect, which will be described in the next section. We then generalize the common features of these phenomena to discuss potential causes and promising lines of research towards a unified explanation of the SIF effect. 

\section{Examples}

\subsection{Pedestrian evacuation}

Perhaps the first formal study of the SIF effect was related to pedestrian flows~\citep{helbing2000simulating}. Modelling crowds like self-driven particles with ``social forces'' interacting among them~\citep{helbing1995social,PhysRevLett.84.1240}, it has been shown that when individuals try to evacuate a room too quickly, they lead to intermittent clogging and a reduced outflow as compared to a calmer evacuation. In this context, the SIF effect is also known as ``freezing by heating"~\citep{Stanley2000}. Trying to exit fast makes pedestrians slower, while calmer people manage to exit faster. This has led people to suggest obstacles close to exits, precisely to reduce friction~\citep{helbing2005self,PhysRevE.87.012802}. Counterintuitively, a slowdown of the evacuation can increase the outflow. Also, in a related study of aircraft evaluation, it was found that there is a critical door width which determines whether ``competitive" evacuation will increase or decrease evacuation time~\citep{Kirchner2003689}. In other words, pushy people will evacuate slower if there are narrow doors (SIF), but will evacuate faster if the doors are wide enough (FIF, faster-is-faster).

\subsection{Pedestrians crossing a road} 

Another example concerns mixed pedestrian and vehicle traffic. Imagine pedestrians are trying to cross a road at a location where there is no traffic light and no pedestrian crossing is marked. This is a typical situation along speed-reduced roads (\emph{e.g.} with a speed limit of 30km/h) or in shared spaces for multi-modal use. Pedestrians would cross when the gap between two successive vehicles exceeds a certain critical separation that ensures a safe crossing of the road. However, there are two types of pedestrians: patient and pushy ones. Pushy pedestrians might force a vehicle to slow down while patient pedestrians would not do this, \emph{i.e.} they would wait for a larger gap. Surprisingly, if all pedestrians were of the patient type, on average they would have to wait for a shorter time period~\citep{Jiang2006567}. 

How does this SIF effect come about? When a pushy pedestrian has slowed a vehicle down, other arriving pedestrians will pass the road, too, and it takes a long time until no further pedestrians arrive, and the stopped cars can accelerate again. During the waiting time, however, a long vehicle queue has formed, such that no large enough gap to cross the road occurs between vehicles until the entire vehicle queue has dissolved. As a consequence, pedestrians will have to wait for a long time until they can cross again. Altogether, it is better if pedestrians wait for large enough gaps such that they don't force vehicles to slow down.

\subsection{Vehicle traffic}

SIF effects are also known from vehicle traffic~\citep{HelbingHuberman1998,Helbing:1998,helbing2001traffic,HelbingNagel2004}. Surprisingly, speed limits can sometimes reduce travel times. This is the case, when the traffic density enters the multi-stable (``meta-stable'') regime. Then, traffic flow is sensitive to disruptions and may break down, which causes largely increased travel times. A speed limit can delay the breakdown of fluid traffic flows, because it reduces the variability of vehicle speeds. This homogenization avoids disturbances in the flow, which are big enough to trigger a breakdown (i.e. have a ``super-critical'' amplitude).

If vehicles go fast, the safety distance between vehicles must be increased. Thus, less vehicles will be able to use a road. For example, at 80km/h, a maximum capacity of about 22 vehicles per km per lane is reached before free traffic flow breaks down. At 120km/h, this capacity is reduced to about 15 vehicles per km per lane. Once vehicles slow down due to an increased density, traffic jams will propagate, as a following car tends to brake more than the vehicle ahead. This phase transition of ``stable" to ``unstable" flow in traffic depends on the desired speed. Thus, to maximize flow, the optimal speed of a highway will depend on the current density. However, the maximum flow lies at the tipping point, and thus a small perturbation can trigger stop-and-go waves which can reduce the highway capacity by 30\%.

A similar consideration applies to over-taking maneuvers~\citep{kesting2007general}. Pushy drivers might force cars in the neighboring lane to slow down when changing lanes to overtake another car, while patient drivers would not do this. As a consequence, pushy drivers may cause a disruption of metastable traffic flow, which may trigger a breakdown (``capacity drop''). Consequently, patient drivers will avoid or delay a breakdown of traffic flow, thereby managing to progress faster on average. One may also formulate this in game theoretical terms. When traffic flow is metastable, drivers are faced with a social dilemma situation: choosing a patient behavior will be beneficial for everyone, while pushy behavior will produce small individual advantages at the cost of other drivers. As a consequence, a ``tragedy of the commons'' results: pushy drivers undermine the stability of the metastable traffic flow, causing congestion that forces everyone to spend more time on travel.  

A complementary phenomenon is observed in Braess's paradox~\citep{Braess,BraessParadox}, where adding roads can reduce the flow capacity of a road network.

\subsection{Traffic light control}

The SIF effect is also found in further systems such as urban traffic light control~\citep{Helbing:2009}. Here, a first-come-first-serve approach works only well at low traffic volumes. Otherwise, forcing vehicles to wait for some time can speed up their overall progress. The reason is that this will produce vehicle platoons, such that a green light will efficiently serve many vehicles in a short time period~\citep{Gershenson2005,GershensonRosenblueth:2011,Zubillaga2014Measuring-the-C}. Similarly, it may be better to switch traffic lights less frequently, because switching reduces service times (due to time lost on amber lights). A ``green wave'', \emph{i.e.}, a coordination of vehicle flows such that several successive traffic lights can be passed without stopping, is another good example demonstrating that waiting (at a red light) may be rewarding altogether.

Similarly interesting observations can be made for self-organized traffic light control (``self-control''), which based on decentralized
flow control (``distributed control'')~\citep{LaemmerEtAl2006,Lammer:2008,Lammer:2010,helbing2013economics}. If each intersection strictly minimizes the travel times of all vehicles approaching it, according to the principle of a ``homo economicus'', this can create efficient traffic flows, when these are low or moderate (``invisible hand phenomenon''). However, vehicle queues might get out of hand when the intersection utilization increases. Therefore, it is beneficial to interrupt travel time minimization in order to clear a vehicle queue when it has grown beyond a certain ``critical'' limit. This avoids spillover effects, which would block other intersections and cause a quick spreading of congestion over large parts of a city. Consequently, waiting for a long queue to be cleared can speed up traffic altogether. Putting it differently, other-regarding self-organization can beat the selfish optimization \`{a} la ``homo economicus'', who strictly does the best, but neglects a coordination with neighbors. 

\subsection{Logistics and supply chains}
 
Similar phenomena as in urban traffic flows are found in logistic systems and supply chains~\citep{helbing2005supply,Helbing2006Self-organizati,Seidel2008,Peters2008}. We have studied, for example, a case of harbor logistics using automated guided vehicles for container transport. Our proposal was to reduce the speed of these vehicles. This reduced the required safety distances between vehicles, such that less conflicts of movement occurred, and the automatic guided vehicles had to wait less. In this way, transportation times could be overall reduced, even though movement times obviously increased. 

We made a similar observation in semiconductor production. So-called ``wet benches'' are used to etch structures into silicium wavers, using particular chemical solutions. To achieve good results, the wavers should stay in the chemical baths longer than a minimum and shorter than a maximum time period. Therefore, it might happen that several silicium wavers need to be moved around at about the same time, while a moving gripper, the ``handler'', must make sure to stay within the minimum and maximum times. It turns out that slightly extending the exposure time in the chemical bathes enables much better coordination of the movement processes, thereby reaching a 30 percent higher throughput.

In a third logistics project, the throughput of a packaging plant had to be increased. One of the central production machines of this plant frequently broke down, such that it was operated at full speed whenever it was operating well. However, this filled the buffer of the production plant to an extent that made its operation inefficient. This effect can be understood with queuing theory, according to which cycle times can dramatically increase as the capacity of a buffer is approached.

\subsection{Public transport}

In public transportation systems, it is desirable to have equal headways between vehicles such as buses, \emph{i.e.}, to reach regular time separations between vehicles. However, the equal headway configuration is unstable~\citep{GershensonPineda2009}. Forcing equal headways minimizes waiting times at stations. Nevertheless, the travel time is not independent of the waiting time, as equal headways imply idling or leaving some passengers at stations. This is because there is a different demand for each vehicle at each station. 
Still, self-organization can be used to regulate the headways adaptively~\citep{Gershenson:2011a}. Considering only local information, vehicles are able to respond adaptively to the immediate demand of each station. With this method, there is also a SIF effect, as passengers wait more time at a station, but reach their destination faster once they board a vehicle because there is no idling necessary to maintain equal headways.

\subsection{Social dynamics}

Axelrod~\citep{Axelrod01041997} 
  proposed an interesting model of opinion formation. In this model, agents may change their opinion depending on the opinion of their neighbors. Eventually, the opinions converge to a stable state. However, if agents switch their opinion too fast, this might delay convergence~\citep{PhysRevLett.101.018701,stark2008slower}. Thus, there is a SIF effect because the fastest convergence will not necessarily be obtained with the fastest opinion change. In this model, there is also a phase transition which is probably related to the optimal opinion change rate~\citep{Vilone2002Ordering-phase-}. There is also experimental evidence of the SIF effect in group decisions. While designing new buildings, slowing down the deliberative process of teams accelerates the design and construction of buildings~\citep{Cross2015}.

Extrapolating these results, one may speculate that high-frequency financial trading~\citep{NARANG2013High-Speed-Trad} may also produce a SIF effect, in the sense that trading at the microseconds scale generates price and information fluctuations which could generate market instabilities leading to crashes and slower economic growth~\citep{Easley2011The-Microstruct}.

In combinatorial game theory~\citep{Siegel2013}, sometimes the best possible move (e.g. taking a queen in chess) is not necessarily the best move in the long term. In other words, having the highest possible gain at each move does not give necessarily the best game result~\citep[pp. 163--171]{russell1995artificial}.

\subsection{Ecology}

If a predator consumes its prey too fast, there will be no prey to consume and the predator population will decline. Thus, a ``prudent predator"~\citep{Slobodkin1961,Goodnight2008} will actually spread faster than a greedy one. A similar SIF effect applies to parasite/host relationships, where parasites taking too many resources from their host are causing their own demise~\citep{Dunne2013}. Over long timescales, evolution will favor symbiotic over parasitic relationships, promoting mechanisms for cooperation which can regulate the interaction between different individuals~\citep{Sachs2004,virgo2013positive}.

We can see that the same principle applies to natural resource management, such as fisheries~\citep{Pauly1998Fishing}. If catches are excessive, there will not be enough fishes left to maintain their numbers, and subsequent catches will be poor. It is estimated that apart from its ecological impact, overfishing has left a void of US\$32 billion per year due to reduced catches~\citep{Toppe2012Aquatic-biodive}. However, regulating how much fish is caught per year is complicated. The maximum sustainable yield varies from species to species~\citep{Maunder2002fishingMSY}, so the calculation of the ``optimal" yields per year is not at all a trivial task. 

\subsection{Adaptation}

Evolution, development, and learning can be seen as different types of adaptation, acting at different timescales~\citep{Aguilar2014The-Past-Presen}. Also, adaptation can be seen as a type of search~\citep{Downing2015Intelligence-Em}. In computational searches, it is known that there needs to be a balance between ``exploration" and ``exploitation"~\citep{Blum:2003:MCO}. An algorithm can explore different possible solutions or exploit solutions similar to those already found. Too much exploration or too much exploitation will lead to longer search times. Too much breadth (exploration) will only explore slightly different types of solution, while too much depth (exploitation) might lead to local optima and data overfitting. A key problem is that the precise balance between exploration (diversification) and exploitation (intensification)  depends the precise search space~\citep{wolpert95no,wolpert1997no} and timescale
~\citep{Gershenson:2010b,Watson2011Global-Adaptati}. 

An example of the SIF was described in biological evolution~\citep{Sellis20122011}. Haploid species (with a single copy of their genome, such as bacteria) can adapt faster than diploid species (with two copies of their genome, such as most plants and animals). Still, in a fast-changing environment, haploids adapt ``too fast", i.e. the population loses genome variation, while diploids can maintain a diversity. Having such a diversity, diploids can adapt faster to changes in their environment, as they can begin an evolutionary search from many different states at once.

In principle, it would be desirable to find a solution as fast as possible, exploiting current solutions. Still, as mentioned, this might lead to suboptimality (SIF) in evolving new features, optimizing a multidimensional function, or training a neural network. To be efficient, search should eventually ``slow down", as it is known from ``simulated annealing''. As too much exploration would be suboptimal also, the critical question is how to find the precise balance to speed up search as much as possible. Computationally, it seems that this question is not reducible~\citep{Wolfram2002}, so we can only know \emph{a posteriori} the precise balance for a given problem. Still, finding this balance would be necessary for adiabatic quantum computation~\citep{Farhi2000,aharonov2008adiabatic}, as if the system evolves too fast, the information is destroyed. 

\section{Generalization}

What do all the above examples have in common? They can be described as complex dynamical systems composed of many non-linearly interacting components. In the above cases, the system can have at least two different states: an efficient and an inefficient one. Unfortunately, the efficient state can be unstable, such that the system will tend to end up in the inefficient state. In the case of freeway traffic, for example, it is well known that the most efficient state (with the highest throughput) is unstable, thereby causing the traffic flow to break down sooner or later (``capacity drop''). To avoid the undesired outcome, the system components must stay sufficiently away from the instability point, which requires them to be somewhat slower than they could be, but as a reward they will be able to sustain a relatively high speed for a long time. If they go faster, the efficient state will break down and trigger another one that is typically slower. This situation might be characterized as a ``tragedy of the commons''~\citep{Hardin13121968}. 

Even though it might be counterintuitive, the SIF effect occurs in a broad variety of systems. For practical purposes, many systems have a monotonic relation between ``inputs" and ``outputs". This is true for systems that ``break"~\citep{Ashby1947}. For example, if temperature is increased in a constrained gas with a constant volume, pressure rises. Yet, if temperature increases too much, then the gas container will break, leading to a pressure reduction. Still, without breaking, many physical and non-physical systems have thresholds, where they become unstable and a phase transition to a different systems state occurs. A typical situation of many-particle systems is that they may get overloaded and turn dysfunctional through a cascading effect. 


To reduce the SIF effect we can seek to adjust 
the interactions which cause a reduction in the system performance~\citep{Gershenson:2010a}. The vehicle traffic case offers an interesting example: when vehicles go too fast (and their density crosses a critical density), their changes in speed will affect other vehicles, generating an amplification of oscillations, which lead to stop-and-go traffic and, as a consequence, to a reduced average speed. If vehicles go slower, then such oscillations can be avoided and the average speed will be higher. The key here is that the critical speed where traffic flow changes from ``laminar" (where FIF) to ``unstable" (where SIF) changes with the density. However, suitably designed adaptive systems, such as driver assistance systems, can be used to drive systems towards their best possible performance in their respective context~\citep{GershensonDCSOS,helbing2015thinking}.

\section{Discussion}

It could be argued that the SIF effect is overly simplistic, as there is only the requirement of having two dynamical phases, where one comes with a reduced efficiency after crossing the phase transition point. Still, as we have presented, the SIF effect shows up in a variety of interesting phenomena at different scales. Thus, we can say that having a better understanding of the SIF effect can be useful and potentially have a broad impact. A challenge lies in characterizing the nature of the different types of interactions which will reduce efficiency~\citep{Gershenson:2011e}.

We can identify the following necessary conditions for the SIF effect:
\begin{enumerate}
\item There is an instability (internal or external) in the system. 
\item The instability is amplified, sometimes through cascading effects.
\item There is a transition from the unstable to a new stable state which leads to inefficiency. 
\item Such a state can be characterized as ``overloaded".
\end{enumerate}

It is worth noting that in some cases, single variables may be stable to perturbations, but their interactions are the ones that trigger instability. This implies that the SIF in these cases has to be studied at two scales: the scale of the components and the scale of the system, as studying components in isolation will not provide enough information to reproduce the SIF effect.

Whether all phenomena with a SIF effect can be described with the same mathematical framework remains to be seen. We believe this is an avenue of research worth pursuing and with relevant implications for the understanding of complex systems.

\section*{Acknowledgments}

We should like to thank Luis \'Alvarez de Icaza, Jeni Cross, Tom Froese, Marios Kyriazis, Gleb Oshanin, Sui Phang, Frank Schweitzer, Diamantis Sellis, Simone Severini, Thomas Wisdom, H\'ector Zenil, and two anonymous referees for useful comments. C.G. was supported by CONACYT projects 212802, 221341, and SNI membership 47907. D.H. was supported by ERC Advanced Grant MOMENTUM 324247.
\bibliographystyle{cgg}
\bibliography{carlos,traffic,eco,complex,evolution,information,computing,sos}

\end{document}